\documentclass{amsart}
\date{July 1, 2003}
\usepackage{amsmath,amsfonts}

\newtheorem{lemma}{Lemma}

\newtheorem{theorem}{Theorem}

\theoremstyle{definition}
\newtheorem{remark}{Remark}

\newcommand{\bk}{{k}}
\newcommand{\bp}{{p}}
\newcommand{\bq}{{q}}

\newcommand{\bx}{{x}}

\newcommand{\gH}{\mathfrak{H}}
\newcommand{\gS}{\mathfrak{S}}
\newcommand{\F}{\mathcal{F}}

\newcommand{\rc}{\mathrm{c}}

\newcommand{\cz}{\mathbb{C}} 
\newcommand{\rz}{\mathbb{R}} 

\newcommand{\Qv}{{Q^\mathrm{\lambda\varphi}}}
\newcommand{\Qbl}{{Q^\mathrm{\bar\lambda\varphi}}}
\newcommand{\dK}{{\int_{-\infty}^{\infty}d\eta}}
\newcommand{\ide}{\frac 1{D^0 + i \eta}}

\newcommand{\idge}{\frac 1{D^0 -\gamma+ i \eta}}
\newcommand{\idvge}{\frac 1{\dv -\gamma + i \eta}}
\newcommand{\idke}{\frac 1{\de -\gamma + i \eta}}
\newcommand{\rl}{\rho^{\lambda}_{{\mathrm{vac}}}}

\newcommand{\dv}{D^{\lambda\varphi}}
\newcommand{\de}{D^{\lambda K^\eps}}
\newcommand{\la}{\langle}
\newcommand{\ra}{\rangle}
\newcommand{\Pgn}{P_{\gamma}^\mathrm{\lambda\varphi}}
\newcommand{\Pgnm}{P_{\gamma}^\mathrm{\mu\varphi}}
\newcommand{\Pvn}{P_-^\mathrm{\lambda\varphi}}
\newcommand{\Pgnbl}{P_{\gamma}^\mathrm{\bl\varphi}}
\newcommand{\Pgnlp}{P_{\gamma}^\mathrm{\lambda'\varphi}}
\newcommand{\Pgpnlp}{P_{\gamma'}^\mathrm{\lambda'\varphi}}
\newcommand{\Pgpo}{P_{\gamma'}^{0}}
\newcommand{\Pgo}{P_{\gamma}^{0}}
\newcommand{\Pon}{P_-^{0}}
\newcommand{\Pge}{P_{\gamma}^\mathrm{\lambda K^\eps}}

\newcommand{\Pgebl}{P_{\gamma}^\mathrm{\bl K^\eps}}
\newcommand{\Pgelp}{P_{\gamma}^\mathrm{\lambda' K^\eps}}
\newcommand{\Pgpelp}{P_{\gamma'}^\mathrm{\lambda'K^\eps}}

\newcommand{\alp}{\boldsymbol{\alpha}}
\newcommand{\ind}{\mathrm{ind}}
\renewcommand{\dim}{\mathrm{dim}}
\newcommand{\Ker}{\mathrm{Ker}}
\newcommand{\Ran}{\mathrm{Ran}}
\newcommand{\vp}{\varphi}
\newcommand{\bl}{\bar \lambda}
\newcommand{\lp}{{\lambda'}}
\newcommand{\eps}{{\varepsilon}}
\newcommand{\qdr}{{Q_3^{\eps}}}
\newcommand{\qvi}{{Q_4^{\lambda,\eps}}}

\def\tr{\mathop{\rm tr}\nolimits} 
\def\Tr{{\rm tr}_{\cz^4}}

\title[Electron-positron pair production]{On the vacuum polarization density caused by an external field}

\thanks{  The author
    acknowledges support through the European Union's IHP
    network Analysis \& Quantum HPRN-CT-2002-00277. He thanks H. Kalf and E. S\'er\'e
for valuable explanations and H. Siedentop for continuing support, advises, and warm hospitality at 
    the LMU-Munich. Furthermore he thanks N. Szpak, V. Shabaev, R. Frank,  and R. Seiringer
    for important comments}

\author[Christian Hainzl]{Christian Hainzl}
\address{CEREMADE, Universit\'e
  Paris-Dauphine, Mar\'echal de Tassigny, F-75775 Paris \& Laboratoire de Math\'ematiques
Paris-Sud-Bat 425, F-91405 Orsay Cedex} \email{hainzl@ceremade.dauphine.fr}

\begin{document}

\markboth{Christian Hainzl}{Electron-positron pair production}

\begin{abstract}
We consider an external potential, $-\lambda \varphi$,
due to one or more nuclei. Following the Dirac picture
such a potential polarizes the vacuum. The polarization density, $\rho^\lambda_{\mathrm{vac}}$,
as derived in physics literature, after a well known renormalization procedure, depends decisively  on the
strength of $\lambda$. For small $\lambda$, more precisely as long as the lowest eigenvalue, $e_1(\lambda)$, of the 
corresponding Dirac operator stays in the gap of the essential spectrum, the integral over the density 
 $\rho^\lambda_{\mathrm{vac}}$ vanishes. In other words the vacuum stays neutral. But as soon as  
$e_1(\lambda)$
dives into the lower continuum the vacuum gets spontaneously charged with charge  $ 2e$. Global charge conservation 
implies 
that two positrons 
were emitted out of the vacuum, this is,
a large enough external potential can produce electron-positron pairs. 

We give a rigorous proof of that phenomenon.
\end{abstract}

\maketitle

\section{Introduction\label{s1}}

In 1934 Dirac and Heisenberg realized that accepting the Dirac picture 
of electrons filling up the negative energy states, called vacuum, consequently implies that a charged nucleus 
thrown into the vacuum causes a redistribution of the Dirac sea, an effect 
denoted as {\em vacuum polarization}. Uehling and Serber in 1935 \cite{Uehling1935, Serber1935},
long before standard renormalization procedure, demonstrated that such an indicated production of
{\em virtual} electron-positron pairs give rise to a modification of the Coulomb potential 
and thus causes energy shifts of bound electrons.

Concerning the traditional Lamb shift, known as the splitting of the 
$2s_{1/2}$- and $2p_{1/2}$-state in hydrogen, this effect only accounts for about $2.5$ percent.
However  the Uehling potential represents the dominating radiative correction in
muonic atoms which emphasizes the importance of vacuum polarization (VP). 
Notice, whereas interaction with a photon field can be treated non-relativistically
there is {\em no} non-relativistic equivalence for VP. It is a
purely  relativistic effect.

Within the framework of QED, VP is treated 
by means of perturbation theory as developed by Dyson, Feynman, and Schwinger.

Only recently Hainzl and Siedentop demonstrated in \cite{HS} that the effective one-particle Hamiltonian
obtained from VP can be handled non-perturbatively and gives rise to
a self-adjoint operator. The effective potential 
we gain is in fact the same as the physicists obtain
after mass and charge renormalization (neglecting photon terms) and
use to calculate the hyperfine structure of bound states. We refer to 
\cite[Section 4]{MPS} for a nice review concerning the influence of VP
on the Lamb shift of heavy atoms.  

The main goal of the present paper 
is to study the vacuum polarization density caused by an 
external field, i.e., by one or more nuclei.
As foreseen by physicists, e.g., \cite{Greineretal1985,GR}, the behavior of the density turns out to 
depend on the lowest eigenvalue of the corresponding Dirac operator.
As long as this eigenvalue stays isolated the integral over the density
vanishes, that means the vacuum stays neutral.
But as soon as that eigenvalue touches the lower continuum
the vacuum gets spontaneously charged, i.e., an electron, more precisely two
electrons due to degeneracy of the ``ground state'', are trapped
in the vacuum and two positrons are emitted. In other words large fields can produce  electron-positron pairs. Such a situation can be realized by heavy ion collision.

\subsection{Model}

The free Dirac operator is given by
\begin{equation}
  D^0: = \alp \cdot\frac1i\nabla + \beta
\end{equation}
in which $\alp,\beta$ denote the $4\times 4$ Dirac matrices.
The
underlying Hilbert space is given by $\gH = L^2(\Gamma)$ with $\Gamma =
\rz^3\times \{1,2,3,4\}$.
We pick
units in which the electron mass is equal to one. 
We regard the case of one, or more, smeared nuclei with
density $n \in L^1(\rz^3) \cap L^\infty(\rz^3)$, non negative, and assuming
$\int_{\rz^3} n =1$.

We remark that it is an experimental fact that the nucleus cannot
shrink to a point. In fact a point nucleus creates instability if
one includes polarization effects, as shown in \cite[Section 3.5]{HS}.

The corresponding electric potential reads
\begin{equation}
  \varphi =  |\cdot|^{-1} * n.
\end{equation}
and the operator to be studied 
is given by
\begin{equation}
 \dv : = D^0 - \lambda \varphi,
\end{equation}
where $\lambda \geq 0$ is a parameter and
can be thought of as $\alpha Z$, $\alpha $ the fine structure constant,
$e := - \sqrt{\alpha}$ the charge of an electron, and $-Z e$ the charge of the nucleus (nuclei).
In the following  we want to allow any value of $\lambda$.

Due to smearing out the Coulomb singularity
the case of large values of $\lambda$ does not
influence the behavior of the essential 
spectrum as well as the self-adjointness as it would be in the case of the Coulomb potential. The following Lemma
is well known, e.g., Weidmann \cite[Theorem 10.37]{Weid}.

\begin{lemma}
Let $\varphi=|\cdot|^{-1}*n$,  $n \in L^1(\rz^3) \cap L^\infty(\rz^3)$, non negative. Then, $\forall \lambda \geq 0$,
$ \dv  = D^0 - \lambda \varphi$ is self-adjoint with domain $H^1(\Gamma)$
and the essential spectrum of $\dv$ is given by
\begin{equation}
 \sigma_{\mathrm{ess}}(\dv ) = (-\infty,-1]\cup [1,\infty).
\end{equation}
\end{lemma}

Throughout the paper we will denote the spectrum
of $\dv$ by $\sigma(\dv)$ and $e_i(\lambda)$ as the corresponding eigenvalues.

The following is well known:
For fixed $\lambda$ there is an infinite number of eigenvalues
which accumulate at $1$ and each $e_i(\lambda)$ 
depends continuously on $\lambda$. For small values of $\lambda$ 
all eigenvalues stay in the gap $(-1,1)$ of the essential spectrum. 
However, for each $i$ one finds a $\lambda_i$ such that
for $\lim_{\lambda \to \lambda_i} e_i(\lambda) = -1$, i.e., the eigenvalue $e_i(\lambda)$ dives
into the lower continuum.
We do not at all discuss what happens to the eigenvalues after reaching the continuum.
In fact one knows from \cite{BG} that below $-1$ there are no embedded eigenvalues. 
However, the behavior of the eigenvalues after reaching $(-\infty, -1]$ 
won't play any role. Our theorems only depend
on the {\em number of eigenvalues, counting multiplicity, that vanish in  the lower continuum}.
Namely, due to our assumption $\varphi \geq 0$, all eigenvalues are 
monotonously decreasing (this is a consequence of \cite[Theorem XII.13]{RS} and the fact that each eigenvalue
has non-positive derivative). That means they will not reappear after
having reached $-1$. The fact that each eigenvalue reaches $-1$ for a large enough parameter
can be seen by e.g., a Theorem of Dolbeaut-Esteban-S\'er\'e \cite{DES}.

We will see that whenever an eigenvalue dives into
the ``sea of occupied states'', i.e., $(\infty,-1]$, a specific number of
$e^-e^+$ pairs are created depending on the degeneracy of the dived eigenvalue.

\subsection{Vacuum polarization density}

As already mentioned above, according to Dirac the vacuum consists of electrons occupying the negative energy states of the
free Dirac operator. If one puts a nucleus into the vacuum,
then the electrons rearrange and one ends up with virtual electron-positron pairs. In other words
the vacuum gets polarized, see e.g., \cite[ page 257]{GR},  for a picture describing this phenomenon,
and \cite{HS} for a ``mathematical'' derivation of the vacuum polarization density,
which follows the idea of the early papers in QED \cite{Dirac1934D,Heisenberg1934,Weisskopf1936,
KrollLamb1949,FrenchWeisskopf1949}. 
For a review about the old fashioned way of QED we refer to \cite{Milonni1994}.

The operator describing this polarization effect is given by
\begin{equation}
\Qv: = \Pvn - \Pon,
\end{equation}
where
\begin{equation}
\Pvn:= \chi_{(-\infty,-1]}(\dv).
\end{equation}
Physically speaking we project onto the occupied states of the Dirac sea.

\begin{remark}
Notice, in the case that the lowest eigenvalue of $\dv$, $e_1(\lambda) $,
is strictly positive, our definition is equivalent to \cite[Equation (12)]{HS},
apart from a minus sign which is chosen to adapt to the definition
in the physics literature.
\end{remark}

Usually the first idea to define a density via $\Qv $ would simply be
taking the diagonal of the Kernel.
Unfortunately, the operator $\Qv$ is not trace class.
The question how to extract from $\Qv$ a physically meaningful
density was first posed in the 30-ies by Dirac \cite{Dirac1934,Dirac1934D} and
Heisenberg \cite{Heisenberg1934} and in more recent literature this procedure is
known as {\it charge renormalization} (see e.g., \cite{FrenchWeisskopf1949,Dyson1949}). As in \cite{HS} we use Cauchy's formula to
express the $\Qv$ in terms of the
respective resolvents (Kato \cite{Kato1966}, Section VI,5, Lemma 5.6)
  \begin{equation}
    \label{eq:12}
    \begin{split}
    \Qv=\Pvn - \Pon&= \frac1{2\pi} \dK
    \left(\frac1{D^0-\gamma+i\eta}-\frac1{\dv-\gamma+i\eta}\right)\\
    &:= \Pgn - \Pgo,
     \end{split}
  \end{equation}
where
  \begin{equation}\label{8}
       \Pgn := \frac 12 -  \frac1{2\pi} \dK
    \frac1{\dv-\gamma+i\eta},
  \end{equation}
with $-1<\gamma<e_j(\lambda)$, $e_j(\lambda)$ being the lowest {\it isolated} eigenvalue of $\dv$.
Notice, that the second equality in \eqref{eq:12} is a consequence of the fact that $\frac1{D^0-z}$
is holomorphic with respect to $z$  in the complex strip between $(-1,1)$ and $\frac1{\dv-z}$
between $(-1,e_j(\lambda))$.

We decompose $\Qv$ into $4$ terms:
\begin{equation}
    \Qv= \lambda Q_1+ \lambda^2 Q_2 + \lambda^3 Q_3 +  \lambda^4 Q^{\lambda}_4,
  \end{equation}
where
\begin{equation}
  \label{eq:13}
  \begin{split}
    Q_1 &:= \frac1{2\pi} \dK \ide \varphi\ide,\\
    Q_2 &:= \frac1{2\pi} \dK \ide \varphi\ide \varphi \ide,\\
    Q_3 &:= \frac1{2\pi} \dK \ide \varphi\ide \varphi
    \ide \varphi \ide,\\
    Q^\lambda_4 &:=\\
  \frac1{2\pi}& \dK \idge \varphi\idge \varphi
    \idvge \varphi \idge \varphi \idge.
\end{split}
\end{equation}
The first three terms we consider by means of its Fourier
representation. A simple variable transform $i\eta \to i\eta + \gamma$
does not change the Kernel of the operators $\hat Q_1 $ to $\hat Q_3$ which is the reason why
we suppressed the $\gamma$ in the denominator. 
The first term
is treated in detail in \cite[Section 3.2]{HS}. There, by a well known renormalization
procedure following Weisskopf \cite{Weisskopf1936} and Pauli and Rose \cite{PauliRose1936}, 
we extracted the corresponding  physical density
\begin{equation}\label{d1}
\rho_1^{\lambda}(\bx) :=e\lambda \F^{-1}\left[\frac{4\pi \hat n(k) C(k)}{|k|^2}\right](x),
\end{equation}
where (see \cite[Equation (21)]{HS})
\begin{equation}
  \label{eq:C}
  \begin{split}
    C(\bk)/k^2 &= \frac12 \int_0^1 dx (1-x^2)
    \log[1+\bk^2(1-x^2)/4] \\
    &= \frac13 \left[(1- \frac 2{\bk^2})\sqrt{1+\frac4{\bk^2}}\log
      \frac{\sqrt{1 + 4/\bk^2} +1}{\sqrt{1 + 4/\bk^2}
        -1}+\frac4{\bk^2}-\frac53\right],
\end{split}
\end{equation}
which was first explicitly written down by Uehling \cite{Uehling1935} and Serber \cite{Serber1935} and later 
by Schwinger \cite{Schwinger1949} and others (see also
\cite{JauchRohrlich1955,KlausScharf1977V, Greineretal1985}). Observe in
\cite[Eq. (52)]{HS} that renormalization consists of subtracting
an operator with infinite diagonal from $Q_1$.
From what rests one defines, in \eqref{d1}, the diagonal 
corresponding to  $Q_1$.  This
subtraction reflects the main difficulty concerning the proof of
our main Theorem.

The second and third term in  \eqref{eq:13} have a well defined
integrable diagonal when
using the Fourier representation. Additionally the density corresponding to $Q_2$
vanishes, either through integration over $\eta$ or due to the fact that
the Dirac matrices are traceless. Quite generally, if we expand $\Tr Q_4^{\lambda}$
into an infinite sum, each term with an even number of $\varphi$
vanishes.

The density corresponding to $Q_3$ is given by
\begin{equation}
  \label{d3}
  \rho_3^\lambda(\bx):
  = e\lambda^3 (2\pi)^{-3}\int_{\rz^3} d\bp  \int_{\rz^3} d\bq \sum_{\sigma=1}^4
  e^{i(\bp-\bq)\cdot\bx}\hat Q_3(\bp,\sigma;\bq,\sigma)
\end{equation}
where $\hat Q_3$ denotes the Kernel of the Fourier representation
\begin{multline}
  \label{eq:23}
  \hat Q_3(\bp,\bq) = \frac1{2\pi}\dK\int_{\rz^3} d\bp_1 \int_{\rz^3}d\bp_2
  (D_\bp+i\eta)^{-1} \circ\hat\varphi(\bp-\bp_1)
  \circ(D_{\bp_1}+i\eta)^{-1}\\
  \circ \hat\varphi(\bp_1-\bp_2)\circ(D_{\bp_2}+i\eta)^{-1}
  \circ\hat\varphi(\bp_2-\bq)\circ (D_{\bq}+i\eta)^{-1}
\end{multline}
with $D_r:= \alp\cdot r+\beta$.

The operator $Q_4^\lambda $ will be shown to be trace class in Lemma \ref{lem2},
so we can define $\rho_4^\lambda$ quite general via the
diagonal of $\lambda^4 \Tr Q^\lambda_4$,
\begin{equation}\label{d4}
\rho_4^\lambda(x):=e\lambda^4 \Tr Q^\lambda_4(x,x)
\end{equation}

Therefore the {\it renormalized} density reads
\begin{equation}
 \rl (x): = \rho_1^\lambda(x) + \rho_3^\lambda(x)+ \rho_4^\lambda(x).
\end{equation}

Before formulating our main theorem it is necessary to  introduce the counting function
$d(\lambda)$ which counts the number of eigenvalues which dived in the 
lower continuum for parameters smaller equal $\lambda$.
\begin{equation}\label{defd}
d(\bar \lambda) := \{\# \mbox{eigenvalues, with multiplicity, that reached} \, -1 \, \,
\mbox{for parameters} \, \lambda \leq \bar \lambda\}
\end{equation} 
\begin{theorem}\label{mt}
Let $n \in L^1(\rz^3) \cap  L^\infty(\rz^3)$ and non negative,
$\varphi=n \ast \frac 1{|\cdot|}$. Then
\begin{equation}\label{emt}
  \int_{\rz^3} \rl(x) dx  = e d(\lambda).
\end{equation}
\end{theorem}

Theorem \ref{mt} exactly reflects the picture which is
presented by physicists, e.g., Greiner et al. \cite{GR,Greineretal1985}:

As long as the external potential, respectively $\lambda$, is so weak that
the lowest eigenvalue, $e_1(\lambda)$, of $\dv$ is in the gap $(-1,1)$ the vacuum
stays neutral (and consists only of {\it virtual} electron-positron pairs).
As soon as the lowest eigenvalue dives into the essential spectrum, $(-\infty,-1]$, i.e.,
the sea of occupied states, the  vacuum immediately gets charged
with charge $2e$ (assuming that the ground state energy of $\dv$ is twice degenerate,
due to the spin). This can be interpreted in the following way: when the unoccupied bound state
dives in the sea of occupied states it traps two electrons which stay in the potential well
of the nucleus (nuclei). Due to Dirac's picture
two ``holes'' emerge which are repelled and emitted as positrons out of the vacuum. 
{\it Consequently we end up with
  real  electron-positron, $e^{-}e^{+}$, pairs}.

This effect of spontaneously emitted positrons is verified in experiment by collision of
heavy nuclei, which when approaching each other create an effective field
strong enough to let the lowest eigenvalue dive into the continuum (see \cite{RMG}).

\begin{remark}
In more recent physics literature, compare e.g., \cite[Equation (7.23)]{GR} or \cite[Equation (230)]{MPS}, the VP-density is ``formally''
denoted as the diagonal of the operator
\begin{equation}\label{dreck}
\frac e2\Tr[\Pvn - P_+^{\mathrm{\lambda \varphi}}],
\end{equation}
with $ P_+^{\mathrm{\lambda \varphi}}:= 1 - \Pvn $. Since
$\Tr[P_+^0 - \Pon]=0$ and  $- P_+^{\mathrm{\lambda \varphi}} + P_+^0 = \Pvn - \Pon $ we see that 
\eqref{dreck}
coincides with our initial  operator  $e  \Tr[\Pvn - \Pon]$. 
\end{remark}

The proof of Theorem \ref{mt} will mainly be
based on two ingredients: A work of Avron, Seiler, and Simon \cite{ASS} concerning the
index of pairs of projectors (see also \cite{Eff}) and arguments of Kato \cite{Kato1966}.

The proof of Theorem \ref{mt} will be given in Section \ref{pmt}. In Section \ref{ind}
we show that for $\tr[\Pvn - \Pon]^{2m+1}$, $m\geq 1$, a result similar
to \eqref{emt} holds.

\section{Result on $\tr[\Pvn - \Pon]^{2m+1}$, with $m \geq 1$}

\label{ind}

Recall that the vacuum polarization is in fact described by the
operator $\Qv=\Pvn - \Pon$. Renormalization is inevitable, since
that operator is not trace class. Nevertheless, due to Klaus and
Scharf \cite{KlausScharf1977T} it is at least an Hilbert-Schmidt
operator. Due to \cite{ASS} (in fact this follows already from Effros \cite{Eff}) 
the traces of
$\left(\Qv\right)^{2m+1}$, $m \geq 1$, are equal. Therefore it is self-evident
to ask for their behavior.

\begin{theorem}\label{st}
Let $n \in L^1(\rz^3) \cap  L^\infty(\rz^3)$, non negative,
$\varphi=n \ast \frac 1{|\cdot|}$. Then, $\forall m \geq 1$,
\begin{equation}
\tr[\Pvn - \Pon]^{2m+1} =  d(\lambda),
\end{equation}
where  $d(\lambda)$ is defined as in \eqref{defd}.
\end{theorem}
\begin{proof}
Notice that, since $\hat \varphi(k) = \hat n(k)\frac {4\pi}{k^2}$,
\begin{equation}
\int_{\rz^3} dk \frac{k^2 \log(2 + |k|) |\hat\varphi(k)|^2}{1+|k|}  \leq
\||\hat n|^2\|_p \left\|\frac{\log(2+|\cdot|)}{|\cdot|^{2}(1+|\cdot|)}\right\|_q.
\end{equation}
Take $q= \frac43$, $p=4$, then the second term on the right hand
side is finite, as well as by Hausdorff-Young inequality
\begin{equation}
\||\hat n|^2\|_4 = \|\hat n\|^2_8 \leq C^3_{8/7} \| n \|^2_{8/7} <
\infty.
\end{equation}
Therefore  the potential
$\varphi$ is regular in the sense of Klaus and Scharf
\cite{KlausScharf1977T}, cf. \cite[Equation (1.7)]{NS}, namely the operator $\Qv\in
\mathfrak{S_2}(\mathfrak{H})$, i.e., $\Qv$ is an Hilbert-Schmidt
operator. Consequently  $\Qv\in
\mathfrak{S_m}(\mathfrak{H})$ for any $m \geq 2$.

To prove the Theorem we first look at the set of all $\lambda \geq
0$ such that the lowest eigenvalue, $e_1(\lambda)$, corresponding
to $\dv$ fulfills
\begin{equation}\label{l1}
e_1(\lambda) > -1.
\end{equation}
This is an open set so that we can  always find a $\gamma$, with
$-1 < \gamma <e_1(\lambda)$ and
\begin{equation}
\Qv= \frac1{2\pi} \dK
    \left(\frac1{D^0-\gamma+i\eta}-\frac1{\dv-\gamma+i\eta}\right)
    = \Pgn - \Pgo
\end{equation} 
We are
going  to show that for $m\geq 1$
\begin{equation}\label{eqw}
\tr[\Pgn - \Pgo]^{2m+1} = 0
\end{equation}
on the set $\{\lambda | e_1(\lambda) >\gamma\}$. Since 
$\gamma$ can be chosen arbitrarily close to $-1$
this infers that 
\begin{equation}
\tr[\Pvn - \Pon]^{2m+1} = 0
\end{equation}
on $\{\lambda | e_1(\lambda) > -1\}$.

To this aim we recall some results from Avron, Seiler, and Simon
\cite{ASS} (see also \cite{Eff}) concerning the index of pairs of
projections:

Regard the family of orthogonal projections $\Pgn$, $\lambda \geq
0$. Since $\Pgn - \Pgnm \in \gS_2(\gH)$, \cite[Proposition
3.2]{ASS} implies that all pairs $(\Pgn, \Pgnm)$ are Fredholm.
Combining \cite[Theorem 3.1]{ASS} and \cite[Theorem 4.1]{ASS} we obtain that for
$m,l\geq 1$
\begin{equation}\label{fred}
\begin{split}
\tr\left[\Pgn - \Pgnm\right]^{2m+1} = &
\tr\left[\Pgn - \Pgnm\right]^{2l+1} = \ind(\Pgn,
\Pgnm)\\ =&
\dim(\Ker \Pgnm \cap \Ran \Pgn) - \dim(\Ker \Pgn \cap \Ran \Pgnm)
\end{split}
\end{equation}
is an integer.
     
\begin{remark}
More generally, a pair $(P,Q)$ of orthogonal
projections is called {\it Fredholm}, if the operator
$T=QP$, as an operator from $\Ran P \to \Ran Q$, is Fredholm.
The corresponding index $\ind (P,Q)$ is defined as
\begin{equation}
\ind (P,Q) : = \ind\, T = \dim(\Ker T) - \dim (\Ran T)^{\perp}.
\end{equation}
\end{remark}
Next we come back to the proof of \eqref{eqw}. 
Observe that on  $\{\lambda | e_1(\lambda) >\gamma\}$
$\Pgn$ is a continuous family with respect to the operator norm.
Namely, using \eqref{8} and expanding the resolvent we get
\begin{multline}\label{rhs}
\|\Pgn - \Pgnm\| \leq |\lambda -\mu|\|\varphi \|\dK \| \frac 1{D^{\mu \varphi} - \gamma + i\eta}\|  
\|\idvge\|\\
\leq | \lambda -\mu| \|\varphi\|\dK \frac1 {\left(\delta_1^2 + \eta^2\right)^{1/2}
\left(\delta_2^2 + \eta^2\right)^{1/2}},
\end{multline}
with $\delta_1 := \min\{\gamma +1, e_1(\mu) - \gamma\}$ and 
$\delta_2 := \min\{\gamma +1, e_1(\lambda) - \gamma\}$. 
Notice, that since $e_1(\lambda)$ is continuous the integral in
the right hand side of \eqref{rhs} can be bounded uniformly
on a small enough closed neighborhood of each $\lambda$ in $\{\lambda |
e_1(\lambda) > \gamma\}$.

Due to \eqref{fred}, $\| \Pgn - \Pgnm\| <1 $ implies $\ind (\Pgn,\Pgnm) = 0$.
Using \cite[Theorem 3.4 (c)]{ASS},
\begin{equation}
\ind (\Pgn,\Pgo)= \ind (\Pgn,\Pgnm) + \ind (\Pgnm,\Pgo),
\end{equation}
the continuity of $\Pgn$  immediately gives that $\ind(\Pgn,\Pgo)=0$ on the whole set
$\{\lambda | e_1(\lambda) >\gamma\}$. Together with \eqref{fred} we arrive at \eqref{eqw}.

Summarizing, the argument given above was based on the fact that on the set  $\{\lambda |
e_1(\lambda) > -1\}$, $\Pgn$ can be {\em continuously deformed} into
$\Pgo$. Throughout the rest of the paper we will
repeat this argument several times.

In the following we consider the case that an eigenvalue has dived
into the lower continuum. We know that there are no
eigenvalues below $-1$. However, for notational simplification
we treat them as if they stay embedded.

Fix now $\bl$ such that $e_1(\bl) \leq -1$ and $e_2(\bl) > -1$,
and $\gamma$ with $-1<\gamma<e_2(\bl)$.
Additionally we choose a $\lp < \bl$ such that
$-1<e_1(\lp)<\gamma$ and a $\gamma'$ with $-1<\gamma' <e_1(\lp)$.
We know
\begin{equation}
\tr\left(\Qbl\right)^{2m+1}= \tr[\Pgnbl -
\Pgo]^{2m+1}=
\ind(\Pgnbl,\Pgo).
\end{equation}
Due to \cite[Theorem 3.4 (c)]{ASS}
\begin{equation}\label{indt}
\ind(\Pgnbl,\Pgo) = \ind (\Pgnbl,\Pgnlp) +
\ind(\Pgnlp,\Pgpnlp)+\ind(\Pgpnlp,\Pgpo).
\end{equation}
The first and third term in the right hand side in \eqref{indt}
vanish which can be seen by repeating the  argument given
above. Namely due to our choice of parameters
$\Pgnbl$ can be continuously deformed into $\Pgnlp$. As well $\Pgpnlp$ can be
continuously deformed into $P^0_{\gamma'}$, which equals $\Pgo$.

Concerning the second term in the right hand side of \eqref{indt}
we note that by Cauchy's formula we obtain
\begin{equation}
\Pgnlp - \Pgpnlp = P_{e_1(\lp)}
\end{equation}
where $P_{e_1(\lp)}$ is the projector on the 
eigenspace corresponding to the eigenvalue $e_1(\lp)$.
Consequently
\begin{equation}
\ind(\Pgnlp,\Pgpnlp)=\tr[P_{e_1(\lp)}].
\end{equation}
By means of our definition \eqref{defd} of $d(\lambda)$,
obviously $\tr[P_{e_1(\lp)}]=d(\bl)$,
whence
\begin{equation}
\tr\left(\Qbl\right)^{2m+1}= d(\bl).
\end{equation}
Repeating this argument whenever an eigenvalue 
dives into the lower continuum, $(-\infty,-1]$, we arrive at the
statement of the theorem.

Notice, due to continuity in $\lambda$ the argument works no matter
how many eigenvalues ``meet'' at $-1$.
\end{proof}

\section{Proof of Theorem \ref{mt}}\label{pmt}

Summarizing the proof of Theorem \ref{st},
we exploited the fact that $\Pvn$ build a continuous family 
of projectors on the non connected intervals
\begin{equation}
[0,\lambda_1) \cup (\lambda_1,\lambda_2)\cup \dots \cup (\lambda_i,\lambda_{i+1}) \dots
\end{equation}
where $\lambda_i$ denotes parameters where an eigenvalue
reaches $-1$. 
As long as $\lambda, \mu$ belong to a connected interval
the index of the corresponding projection vanishes,
\begin{equation}
\ind(\Pvn, {P_-^\mathrm{\mu\varphi}}) =0,
\end{equation}
but if $\lambda$ moves to a different not connected interval 
the index jumps by an integer value.

In order to prove Theorem \ref{mt} we first recall
the definition of the density
\begin{equation}
 \rl (x) = \rho_1^\lambda(x) + \rho_3^\lambda(x)+ \rho_4^\lambda(x),
\end{equation}
the terms on the right hand side being defined in \eqref{d1},
\eqref{d3}, \eqref{d4}. By means of our explicit choice of
$\rho_1^\lambda$ via Fourier transform $\hat \rho_1^\lambda(k) =e \lambda 4\pi \hat
n \frac{C(k)}{k^2}$ and the fact that $\lim_{|k| \to 0}\frac{C(k)}{k^2}=0$
we immediately obtain
\begin{equation}
\int_{\rz^3} \rho_1^\lambda(x) dx =\hat \rho_1^\lambda(0) =0.
\end{equation}
Therefore, our goal in the following will be to show
that for all $\lambda$
\begin{equation}
\int_{\rz^3} \rho_3^\lambda(x) dx =0, \qquad \int_{\rz^3} \rho_4^\lambda(x) dx =
e d(\lambda). 
\end{equation}

It still remains to show that $Q^\lambda_4$ is trace class
which works analogously to \cite[Lemma 3]{HS}.

\begin{lemma}\label{l1t}
\begin{equation}\label{41}
\tr|Q^\lambda_4| = \|Q^\lambda_4\|_1 \leq C^\mu \|\varphi\|_4^4,
\end{equation}
with an appropriate constant $C^\mu$ depending on $\mu:=\min\{\gamma +1, e_i(\lambda) - \gamma\}$,
$ e_i(\lambda)$ denoting the lowest isolated eigenvalue of $\dv$.
\end{lemma}
\begin{proof}
Let $e_i(\lambda)$ be the lowest isolated eigenvalue of
$\dv$, then as usually, we choose a $\gamma$ with $-1 <\gamma < e_i(\lambda)$.
Using \eqref{eq:13} we obtain (apart from a factor $\frac1{2\pi} $)
\begin{multline}\label{sac}
\|Q^\lambda_4\|_1 \leq \\
 \dK\left\| \idge \varphi\idge
    \varphi \idvge \varphi \idge \varphi
    \idge\right\|_1 \\
\leq  \dK\left\| \idge \varphi\idge
    \varphi \idge \varphi \idge \varphi
    \idge\right\|_1 \\ \times \left\|(D^0 -\gamma +  i\eta) 
\frac {1}{\dv - \gamma + i \eta }\right\|, 
\end{multline}
with $\|(D^0 -\gamma +  i\eta) \frac {1}{\dv - \gamma + i \eta }\| \leq 1 + \lambda \|\varphi 
\frac {1}{\dv - \gamma  }\|$ which depends on $\mu$.
Moreover, with $\|(D^0 +  i\eta) \frac {1}{D^0 - \gamma + i \eta }\| \leq 1 +
\mu^{-1}$,
\begin{multline}\label{fac}
 \dK\left\| \idge \varphi\idge
    \varphi \idge \varphi \idge \varphi
    \idge\right\|_1 \\
\leq \dK\left\| \ide \varphi\ide
    \varphi \ide \varphi \ide \varphi
    \ide\right\|_1(1 +
\mu^{-1})^{5} \\ \leq \dK \left\| \varphi \ide
  \right\|_4^3 \left\| \varphi \ide \ide
  \right\|_4(1 +
\mu^{-1})^{5} .
\end{multline}
Applying an inequality of Simon
\cite[Theorem 4.1]{Simon1979T},
\begin{equation}
\label{eq:18}
\|f(x)g(-i\nabla)\|_4 \leq (2\pi)^{-3/4}
\|f\|_4 \|g\|_4,
\end{equation}
to the factors in \eqref{fac}, gives
\begin{equation}
  \label{eq:14}
\begin{split}
  \|\varphi\ide\|_4 & \leq \frac{1}{2^{1/4}\pi^{3/4}}
\|\varphi\|_4\|1/\sqrt{|\cdot|^2+1+\eta^2}\|_4\\
 \|\varphi\ide\ide\|_4 & \leq \frac{1}{2^{1/4}\pi^{3/4}}
\|\varphi\|_4\|1/({|\cdot|^2+1+\eta^2})\|_4.
\end{split}
\end{equation}
Putting all together and evaluating the integrals (cf. \cite[Lemma 3]{HS})
we arrive at 
\begin{equation}
\|Q^\lambda_4\|_1 \leq C^\mu \|\varphi\|^4_4,
\end{equation}
with an appropriate $C^\mu$.
\end{proof}

In the following we will proceed analogously to 
the proof of Theorem \ref{st}. We will  circumvent the problem that
$\Pvn - \Pon$ is not trace class
by defining a family of trace class operators
$K^\eps$ converging strongly to $\varphi$.
We define $K^\eps$ via its Fourier representation
\begin{equation}\label{42}
\hat K^\eps(p,q): = f_\eps(p) \hat \varphi_\eps(p-q)f_\eps(q),
\end{equation}
with
\begin{equation}\label{43}
 f_\eps(p): = \chi(1/\eps -  |p|), \qquad   \varphi_\eps(x):= \varphi(x)\chi(1/\eps - |x|),
\end{equation}
$\chi$ denoting the Heaviside step function.
Obviously $f_\eps \to 1$ and $\varphi_\eps \to \varphi$ pointwise when $\eps \to 0$.

The family of operators
\begin{equation}
\de: = D^0 - \lambda K^\eps
\end{equation}
turn out to converge strongly to $\dv$.
For convenience we define  $Q^{\lambda\mathrm{K^\eps}}$ via an appropriate $\gamma$
chosen corresponding to $\dv$,
\begin{equation}\label{mou}
Q^{\lambda\mathrm{K^\eps}} := \Pge - \Pgo,
\end{equation}
leaving away the subscript $\gamma$, since it will not cause ambiguities.
 Furthermore $\Pge - \Pgo$
will be trace class so we can repeat the arguments
given in the proof of Theorem \ref{st}. Since we already removed the ``bad''
part of $\Pvn - \Pon$ by charge renormalization (in \cite{HS}), that is the part
of $Q_1$ which prevents  $\Pge - \Pgo$ from being trace class,
it suffices to show that
$\qdr$ (respectively $\qvi$) converge (in trace norm) to $Q_3$ (respectively $
Q_4^\lambda$).
$\qdr$ and  $\qvi$ are  terms we obtain by expanding \eqref{mou}.

Notice, due to definition \eqref{d3},
$\int \rho_3^\lambda(x) dx = e\lambda^3 \int_{\rz^3}  \Tr \hat Q_3(p,p)dp$.

First we state a few useful properties of $K^\eps$.
\begin{lemma}\label{lem2}
{\rm{(a)}} \,\, For all $\eps > 0$, $K^\eps$ is trace class
and $K^\eps\geq 0$. Moreover,
$\sigma_{\mathrm{ess}}(\de) = (-\infty,-1] \cup [1,\infty)$.\\
{\rm{(b)}} \,\,  $\de \to \dv $ strongly as $\eps \to 0$. \\
{\rm{(c)}} \,\, $\Pge - \Pgo$ is trace class for all  $\eps > 0$ if
$\gamma \not \in \sigma (\de)$.
\end{lemma}
\begin{proof}
(a) \, \, The fact that $K^\eps $ is trace class
is a direct consequence of Lemma \ref{lapp}. In Fourier
representation we can decompose 
\begin{equation}\label{pos}
\hat K^\eps(p,q) = L_\eps^* L_\eps (p,q),
\end{equation} 
with $L_\eps(p,p') = f_\eps(p) h_\eps(p-p')$, $h_\eps(p) =(2\pi)^{-3/4} \widehat {\sqrt{\varphi_\eps}}(p)$.
By our choice of $f_\eps$ and $\varphi_\eps$
\begin{equation}
\int_{\rz^3} \int_{\rz^3}| L_\eps(p,q)|^2 dp dq < \infty
\end{equation} 
whence $K^\eps$ is trace class. Equation \eqref{pos} immediately implies
$K^\eps \geq 0$. The compactness of $K^\eps$ yields  
$\sigma_{\mathrm{ess}}(\de) = (-\infty,-1] \cup [1,\infty)$
by Weyl's Theorem.

\noindent
(b) \,\, For $\psi \in H^1(\Gamma)$
\begin{equation}
\lim_{\eps \to 0} \| [\de - \dv]\psi\| = \lambda\lim_{\eps \to 0} \| [K^\eps - \varphi]\psi\| =0,
\end{equation}
since ${{\hat { K^\eps}}}^* \hat K^\eps \to \hat \varphi \ast \hat \varphi$ in the sense of
distributions, and these operators are bounded.

\noindent
(c) \,\, Let $e^\eps_i(\lambda)$ be the lowest isolated eigenvalue 
of $\de$. Then, with $-1 < \gamma < e^\eps_i(\lambda)$,
\begin{multline}
 \Pge - \Pgo= \frac 1{2\pi}\dK \left( \idge -\idke \right)\\
= \lambda \frac 1{2\pi} \dK \idge K^\eps \idke .
\end{multline}
Consequently
\begin{equation}
\|\Pge - \Pgo\|_1  \leq
\lambda \|K^\eps\|_1 \frac 1{2\pi} \dK \frac 1{\left((1+\gamma)^2 + \eta^2\right)^{1/2}
\left(\bar \delta^2 + \eta^2\right)^{1/2}},
\end{equation}
with $\bar \delta:=\min\{ \gamma + 1, e^\eps_i(\lambda) - \gamma\}$
using $\|\idke\| \leq ({\bar\delta}^2 + \eta^2)^{-1/2}$ and  $\|\idge\| 
\leq \left((1+\gamma)^2 + \eta^2\right)^{-1/2}$.
\end{proof}

Let us fix an arbitrary $\lambda$ such that
$e_1(\lambda) > -1$ and  $\gamma$ with $-1<\gamma < e_1(\lambda)$. Since
$\de \to \dv$ strongly, a Theorem of Kato \cite[VIII-5, Theorem 5.1]{Kato1966} tells us that
$\sigma(\de)$ is asymptotically concentrated in any open set containing
$\sigma(\dv)$. Thus we can find a $\delta $ small enough such that  $
\gamma < e_1(\lambda) - \delta$ and a corresponding
$\eps_0$ such that for all $\eps \leq \eps_0$,
$\sigma(\de)$ is concentrated in a $\delta$-neighborhood of $\sigma(\dv)$,
in particular $e_i^\eps(\lambda) > e_1(\lambda) - \delta$ for each eigenvalue $e_i^\eps(\lambda)$
of $\de$.

Thus we are able to guarantee that
$\Pge$ can be continuously deformed into $\Pgo$.
Therefore we can argue analogously to the proof of Theorem \ref{st}
combined with the trace class property of $\Pge - \Pgo$ to obtain
\begin{equation}
\tr[\Pge - \Pgo] = \ind (\Pge,\Pgo) = 0.
\end{equation}
Expanding the resolvent this implies
\begin{multline}\label{ja}
0= \lambda \tr \dK \ide K^\eps\ide + 
  \lambda^2 \tr \dK \ide K^\eps\ide K^\eps \ide \\
  + \lambda^3\tr  \dK \ide K^\eps\ide K^\eps\ide K^\eps \ide + \\
 \lambda^4  \tr \dK \idge K^\eps\idge K^\eps \times \\ \times
    \idke K^\eps \idge K^\eps \idge \\
:= \lambda \tr Q_1^\eps +\lambda^2 \tr Q_2^\eps +\lambda^3 \tr Q_3^\eps +
\lambda^4 \tr \qvi .
\end{multline}
Observe, this holds in particular in a small neighborhood of $0$. Thus,
since the fourth term on the right hand side is of order $O(\lambda^4)$,
each term in \eqref{ja} vanishes separately.  In particular these terms
we are interested in:
\begin{equation}\label{58}
\tr Q_3^\eps =0, \quad \tr \qvi = 0,
\end{equation}
the latter one on the set $\{\lambda| e_1(\lambda) >-1\}$.

Assume for a moment we have already shown
\begin{equation}\label{toprove}
e\lambda^3\lim_{\eps \to 0}\tr Q_3^\eps =\int_{\rz^3} \rho^\lambda_3(x) dx , 
\quad e\lambda^4 \lim_{\eps \to 0} \tr \qvi = \int_{\rz^3} \rho^\lambda_4(x) dx,
\end{equation}
then by \eqref{58} obviously $\int_{\rz^3} \rho^\lambda_3(x) dx =0$ and $ \int_{\rz^3} \rho^\lambda_4(x) dx = 0$ 
whence Theorem \ref{mt} on  $\{\lambda| e_1(\lambda) >-1\}$.

In order to prove \eqref{toprove} we formulate an auxiliary Lemma:
\begin{lemma}\label{lem3}
{\rm{(a)}}
There exists a non negative function $g\in L^1(\rz^3)$,
such that 
\begin{equation}
\left| \Tr \hat Q_3^\eps(p,p)\right| \leq g(p)
\end{equation}
uniformly in $\eps$.\\
{\rm{(b)}}
Let $e_i(\lambda)$ be the lowest isolated eigenvalue of $\dv$. Fix $\gamma$ and $\bar \delta$  with  
$-1+\bar \delta < \gamma <e_i(\lambda)-\bar \delta$. Furthermore
fix $\eps_0$ such that for all $\eps \leq \eps_0$,
$\sigma(\de)$ is in a $\bar \delta $-neighborhood of $\sigma(\dv)$. Then
\begin{equation}\label{61}
\|\qvi \|_1 \leq C^\mu \|\varphi\|_4^4
\end{equation}
uniformly in $\eps \leq \eps_0$,
where $C^\mu$ is an appropriate constant depending on 
$\mu := \min\{\gamma + 1 -\bar \delta , e_i(\lambda) - \bar \delta - \gamma\}$,
and
\begin{equation}\label{tracon}
\lim_{\eps\to 0}\tr{\qvi} = \tr Q^\lambda_4.
\end{equation}
\end{lemma}
\begin{proof} (a) \, \,
We will proceed similarly to \cite[Lemma 4]{HS}. For completeness
we will repeat some parts of the proof.
The
``eigenfunctions'' of the free Dirac operator in momentum space are
\begin{equation}
  \label{eq:19}
  u_\tau(\bp):=
  \begin{cases}
    \frac 1 {N_+(\bp)} \begin{pmatrix} \sigma \cdot \bp\, {\bf e}_{\tau}\\
      -(1-E(\bp))\mathbf{e}_\tau\end{pmatrix} & \tau = 1,2,\\
    \frac 1 {N_-(\bp)} \begin{pmatrix} \sigma \cdot \bp \, {\bf e}_{\tau}\\
      -(1+E(\bp))\mathbf{e}_\tau\end{pmatrix} &\tau =3,4
  \end{cases}
\end{equation}
with $ {\bf e}_{\tau} := (1,0)^t$ for $\tau=1,3$ and $ {\bf e}_{\tau}:=
(0,1)^t$ for $\tau=2,4$ and
\begin{equation}
N_+(\bp) = \sqrt{2 E(\bp)(E(\bp)-1)}, \,\,\,\, N_-(\bp) = \sqrt{2 E(\bp)(E(\bp)+1)}.
\end{equation}
The indices $1$ and $2$ refer to positive ``eigenvalue'' $E(\bp)$ and
the indices $3$ and $4$ to negative $-E(\bp)$. Using Plancherel's theorem we get
\begin{multline}
  \label{v32}
\Tr\hat Q_3^\eps (p,p) =  \sum_{\tau_0=1}^4 \la  u_{\tau_0}(p)|\hat  Q_3^\eps| u_{\tau_0}(p)\ra 
  = \frac1{(2\pi)^{4}} \int_{\rz^3}
  d\bp_1\int_{\rz^3}d\bp_2
  \sum_{\tau_0,\tau_1,\tau_2, = 1}^4 \times \\ \times
  \la u_{\tau_0}( \bp)|\hat K^\eps |u_{\tau_1}(\bp_1)\ra\la u_{\tau_1}(\bp_1)
  |\hat K^\eps | u_{\tau_2}(\bp_2)\ra \la u_{\tau_2}(\bp_2) |\hat K^\eps | u_{\tau_0}(\bp)\ra  \\
  \times \dK \frac 1{(ia_{\tau_0} E(\bp) - \eta)(ia_{\tau_1} E(\bp_1)
    - \eta) (ia_{\tau_2} E(\bp_2) - \eta)(ia_{\tau_0} E(\bp) - \eta)},
\end{multline}
with $a_\tau = 1$ for $\tau=1,2$ and $a_\tau =- 1$ for $\tau=3,4$.
The integral over $\eta$ is seen to vanish by Cauchy's theorem, if all four
$a_{\tau_j}$ have the same sign. In fact we only treat one case.
The others then work analogously.

Set
\begin{equation}
  a_{\tau_2} = -1,\, \, \, a_{\tau_0} = a_{\tau_1}= 1.
\end{equation}
Using $f_\eps \leq 1$ the first factor in (\ref{v32}) can be estimated by
\begin{multline}
  \label{v33}
\sum_{\tau_0=1,2} \la u_{\tau_0}( \bp)|\hat K^\eps |u_{\tau_1}(\bp_1)\ra 
\sum_{\tau_1=1,2}  \la u_{\tau_1}(\bp_1)
  |\hat K^\eps | u_{\tau_2}(\bp_2)\ra \sum_{\tau_2=3,4} 
\la u_{\tau_2}(\bp_2) |\hat K^\eps | u_{\tau_0}(\bp)\ra \\
\leq |\hat \varphi (\bp- \bp_1) \hat \varphi(\bp_1 -
  \bp_2)\hat
  \varphi(\bp_2-\bp)|\Big|
  \tr_{\cz^2} \Big[ \frac{\sigma\cdot \bp \sigma\cdot \bp_1 +
    (1-E(\bp))(1-E(\bp_1))}{N_-(\bp_2)^2N_+(\bp)^2N_+(\bp_1)^2}\times \\ \times
  \big[\sigma\cdot \bp_1 \sigma\cdot \bp_2 +
  (1-E(\bp_1))(1+E(\bp_2))\big]  \big[\sigma\cdot \bp_2 \sigma\cdot \bp +
  (1+E(\bp_2))(1-E(\bp))\big]\Big]\Big|\\
\leq \rc |\hat \varphi (\bp- \bp_1) \hat \varphi(\bp_1 -
  \bp_2)\hat
  \varphi(\bp_2-\bp)| \frac{| \bp\cdot \bp_2 - (E(\bp_2)-1)(1+E(\bp))| +
    |\bp\wedge\bp_2|}{N_-(\bp_2)N_+(\bp)}.
\end{multline}
($\rc$ being  a generic constant.)
Since
\begin{multline}
  \frac1{2\pi}\dK \frac 1{(i E(\bp) - \eta)(i E(\bp_1) - \eta)
    (-i E(\bp_2) - \eta)(i E(\bp) - \eta)} \\
  = \frac 1{2(E(\bp) + E(\bp_1))^2
    E(\bp)}
\end{multline}
our term of interest (\ref{v32}) is bounded by a constant times
\begin{multline}\label{v34}
  \int_{\rz^3} d\bp_1\int_{\rz^3}d\bp_2
  |\hat \varphi (\bp- \bp_1) \hat \varphi(\bp_1 -
  \bp_2)\hat
  \varphi(\bp_2-\bp)| \times \\
  \times\frac{| \bp\cdot \bp_2 - (E(\bp_2)-1)(1+E(\bp))| +
    |\bp\wedge\bp_2|}{2N_-(\bp_2)N_+(\bp)(E(\bp) + E(\bp_1))^2
    E(\bp)}.
\end{multline}
Substituting $\bp_2 \to \bp_2 + \bp$,  $\bp_1 \to \bp_1 + p_2 +
\bp$ we get
\begin{multline}\label{v35}
 |\eqref{v34}| \leq  \int_{\rz^3} d\bp_1\int_{\rz^3}d\bp_2
  |\hat \varphi (\bp_1+ \bp_2) \hat \varphi(
  \bp_1)\hat
  \varphi(\bp_2)| \times \\
  \times\frac{| \bp\cdot (p_2 + p) - (E(p_2 + p)-1)(1+E(\bp))| +
    |\bp\wedge(p_2 + p)|}{2N_-(p_2 + p)N_+(\bp)(E(\bp) + E(\bp_1+p_2+p))^2
    E(\bp)}.
\end{multline}
Since 
$$ | \bp\cdot (p_2 + p) - (E(p_2 + p)-1)(1+E(\bp))| +
    |\bp\wedge p_2|\leq 4|p||p_2|$$
we obtain as an upper bound the function 
\begin{equation}
\bar g(p):=  \int_{\rz^3} d\bp_1\int_{\rz^3}d\bp_2
  |\hat \varphi (\bp_1+ \bp_2) \hat \varphi(
  \bp_1)\hat
  \varphi(\bp_2)| |p_2| \frac 1{N_-(p_2 + p)  E(\bp)^3},
\end{equation}
which is obviously in $L^1(\rz^3)$,
whence (a) is proven.

\noindent (b) \,\,
Analogously to \eqref{sac} and \eqref{fac} we get (apart from a constant)
\begin{multline}
\|\qvi\|_1 \leq \dK \times \\ \times
 \left\| \idge K^\eps\idge
     K^\eps \idke  K^\eps \idge  K^\eps
    \idge\right\|_1 \\
\leq \dK\left\| \ide  K^\eps\ide
     K^\eps \ide  K^\eps \ide  K^\eps
    \ide\right\|_1 \times\\ \times \left( 1 + 
\left\| {K^\eps}\frac 1 {\de - \gamma  }\right\|\right)(1 +
\mu^{-1})^{5}.
\end{multline}
The first term in the third line
is trace class, so we can evaluate it in Fourier representation.
Since $f_\eps \leq 1$ we are in the situation of Lemma \ref{l1t}
and end up with
\begin{equation}
\|\qvi\|_1 \leq \rc  \left( 1 + \|K^\eps\|
\mu^{-1} \right)(1 +
\mu^{-1})^{5} \|\varphi\|_4^4
\end{equation}
which implies \eqref{61} since $\|K^\eps\| $ is uniformly bounded.

In order to prove \eqref{tracon} it suffices to
show
$$\|  Q_5^{\lambda,\eps} -  Q_5^{\lambda} \|_1 \to_{\eps \to 0} 0,$$
since $\tr Q_4 = 0$,
with
\begin{equation}
Q_4 = \frac1{2\pi} \dK \ide \varphi\ide \varphi
    \ide \varphi \ide\varphi \ide.
\end{equation}
In fact, due to Lemma \ref{l1t}, $Q_4$ is traceclass and using the Fourier representation
one easily sees that its trace vanishes. Indeed each operator $Q_{2n}$ with an even number
of potentials has vanishing trace, which is well
known as Furry's Theorem.
For convenience we denote
\begin{equation}
\begin{split}
Q_5^{\lambda,\eps} &= \frac1{2\pi} \dK N^\eps_\eta M^\eps_\eta M^\eps_\eta M^\eps_\eta M^\eps_\eta \\
Q_5^{\lambda} & =  \frac1{2\pi} \dK N_\eta M_\eta M_\eta M_\eta M_\eta,
\end{split}
\end{equation}    
where $N^\eps_\eta= \idke K^\eps\idge$, $M^\eps_\eta=K^\eps\idge$, $N_\eta= \idvge \varphi \idge $
and $M_\eta = \varphi \idge$.
A straightforward calculation gives
\begin{multline}\label{suminteta}
\|  Q_5^{\lambda,\eps} - Q_5^{\lambda} \|_1  \leq \sup_\eta \|M^\eps_\eta - M_\eta\|_4 \frac1{2\pi} \dK 
\Big[ \| N^\eps_\eta M^\eps_\eta M^\eps_\eta M^\eps_\eta \|_{4/3}\\ +\| N^\eps_\eta M^\eps_\eta M^\eps_\eta M_\eta \|_{4/3} 
+\| N^\eps_\eta M^\eps_\eta M_\eta M_\eta \|_{4/3} +\| N^\eps_\eta M_\eta M_\eta M_\eta \|_{4/3}\Big]
\\ + \sup_\eta \|N^\eps_\eta - N_\eta\|_\infty \frac1{2\pi} \dK \| M_\eta M_\eta M_\eta M_\eta\|_{1}.
\end{multline}
Notice, since trace ideals fulfill $\gS_{p} \subset \gS_1$, for $p > 1$, we can easily estimate both integrals in \eqref{suminteta}
in analogy to \eqref{61} and \eqref{41}.

Obviously
\begin{equation}
\| M^\eps_\eta - M_\eta \|_4 \leq \Big\| \frac 1{\sqrt{p^2 + 1}}[\hat \vp - \hat K^\eps ]\Big\|_4 := \Big(\int dp dq f_\eps(p,q)\Big)^{1/4}.
\end{equation}
Recalling the definition of $K^\eps$ in \eqref{42},
we see that $f_\eps (p,q) \to_{\eps \to 0} 0$ pointwise,
as well as
$$ f_\eps (p,q) \leq \frac 1{p^2 + 1} \big[|\hat \vp| \ast | \hat \vp|\big]^2(p-q)\frac 1{q^2 + 1} \in L^1 (\rz^3 \times \rz^3),$$
which implies, by the diminated convergence theorem, that
$\sup_\eta \| M^\eps_\eta - M_\eta \|_4 \to_{\eps \to 0} 0 $.

Notice that
\begin{equation}
 \idke \to \idvge \quad \mathrm{strongly}
\end{equation}
and 
\begin{equation}
 K^\eps \idge \to \vp \idge \quad \mathrm{in} \,\, \gS_4,
\end{equation}
both uniformly in $\eta$. Together with the fact that $\vp \idge $
is compact (even in $\gS_4$), i.e., it  can be approximated in norm
by a finite rank operator, we conclude
$$\sup_\eta \|N^\eps_\eta - N_\eta\|_\infty \to_{\eps \to 0} 0 $$
which yields \eqref{61}.
\end{proof}

Now we are ready to prove \eqref{toprove}.
Obviously, due to our definition \eqref{42} and \eqref{43},
\begin{equation}
\Tr \hat Q_3^\eps (p,p) \to \Tr \hat Q_3(p,p)
\end{equation}
pointwise as $\eps \to 0$. By means of Lemma \ref{lem3} (a) and the
dominated convergence theorem
\begin{equation}
e\lambda^3 \lim_{\eps \to 0}\int_{\rz^3}  \Tr \hat Q_3^\eps (p,p)dp = e\lambda^3 
\int_{\rz^3}\Tr \hat Q_3(p,p) dp = \int_{\rz^3} \rho_3^\lambda (x) dx.
\end{equation}

By means of Lemma \ref{lem3} (b) 
we obtain 
\begin{equation}
e\lambda^4 \lim_{\eps \to 0}  \tr \qvi  = e\lambda^4 
\tr Q_4^\lambda = \int_{\rz^3} \rho_4^\lambda(x) dx ,
\end{equation}
whence \eqref{toprove}.

Fix again $\bl$ such that $e_1(\bl) \leq -1$ and $e_2(\bl) > -1$.
(As in the previous section we use the notation $e_1(\bl)$ for convenience.
The argument works whatever happens to the eigenvalue after reaching
the lower continuum. Let us remark, that for our results only the features of the eigenvalues
before ``diving'' play a role.) 
We can find a $\delta>0$ small enough such that the following holds:
We can choose a  $\gamma$ with $-1+\delta <\gamma< e_2(\bl) - \delta$.
Additionally we choose a $\lp<\bl$ such that
$-1+\delta<e_1(\lp)<\gamma-\delta$ and a $\gamma'$ with $-1+\delta<\gamma' <e_1(\lp)-\delta$.
Moreover we find, due to Kato \cite[VII-5, Theorem 5.1]{Kato1966}, an 
$\eps_0$ such that for all $\eps \leq \eps_0$,
$\sigma(D^{\bl\mathrm{K}^\eps})$ is in a $\delta$-neighborhood 
of $\sigma(D^{\bl\mathrm{\varphi}} )$ as well as 
$\sigma(D^{\lp\mathrm{K}^\eps})$ in a $\delta$-neighborhood 
of $\sigma(D^{\lp\mathrm{\varphi}} )$.
We can write
\begin{equation}\label{trs}
\tr[\Pgebl - \Pgo] = \tr[ \Pgebl - \Pgelp] + \tr[  \Pgelp - \Pgpelp] 
+\tr[\Pgpelp -  \Pgo ].
\end{equation}
By our choice of parameters $\Pgebl$ can be continuously deformed into
$\Pgelp$, as well as $\Pgpelp$ into $P^0_{\gamma'} $  which equals $ \Pgo$.
Consequently the first and the third term on the right hand side of \eqref{trs}  vanish.
Due to Cauchy's formula
\begin{equation}
 \tr[  \Pgelp - \Pgpelp] 
= \tr[P_{e_1(\lp)}],
\end{equation}
due to the fact that by our choice of parameters  the  eigenspace of the set 
$\{ e_j^\eps(\lp)| \gamma' < e_j^\eps(\lp) < \gamma\}$
has the same dimension as the eigenspace of $e_1(\lp)$. Recall  $P_{e_1(\lp)}$ denotes the
projector on the eigenspace corresponding to $e_1(\lp)$.
By definition \eqref{defd}
\begin{equation}
\tr[P_{e_1(\lp)}] = d(\bl).
\end{equation}
Whence 
\begin{equation}
\tr[\Pgebl - \Pgo] = d(\bl) .
\end{equation}
Expanding the left hand side as in \eqref{ja} and using the fact that we already know
that  the first three terms vanish
we see
\begin{equation}
{\bl }^4 \tr Q_4^{\bl,\eps} = d(\bl).
\end{equation}
By means of Lemma \ref{lem3} (b)
$ \lim_{\eps \to 0} \tr Q_4^{\bl,\eps} = \tr Q_4^{\bl}$, so
we infer
\begin{equation}
\int_{\rz^3} \rho_4^{\bl}(x) dx =ed(\bl) .
\end{equation}

Repeating this argument whenever an $e_i(\lambda)$ dives into $(-\infty,-1]$
we arrive at the Theorem.

\begin{appendix}
\section{Criterion for a Kernel to be trace class}

It is well known that given an integral operator
via a Kernel $K(x,y)$,
the fact that $\int_{\rz^n} dx K(x,x) < \infty$ does not
at all guarantee that $K$ is trace class.

For a specific class of Kernels we give a sufficient condition
for the corresponding operator to be trace class.

\begin{lemma}\label{lapp}
Let
\begin{equation}
K(x,y) = f_1(x) g(x-y)f_2(y),
\end{equation}
with $f_1, f_2 \in L^2(\rz^n)$ and $\hat g \in L^1(\rz^n)$.
Then $K$ is trace class.
\end{lemma}
\begin{proof}
We can write
\begin{equation}
\hat g(k) = |\hat g (k)|^{1/2}|\hat g (k)|^{1/2} {\mathrm{sgn}} (\hat g(k)).
\end{equation}
Define
\begin{equation}
h_1(x): = (2\pi)^{-3/2}\F^{-1}[ |\hat g|^{1/2}](x), \quad h_2(x) : = \F^{-1}[ |\hat g|^{1/2} {\mathrm{sgn}} (\hat g)](x),
\end{equation}
such that
\begin{equation}
g =  \F^{-1}[ \hat g] = (2\pi)^{-3/2} \F^{-1}[ |\hat g|^{1/2}] \ast \F^{-1}[ |\hat g|^{1/2} {\mathrm{sgn}} (\hat g)] = h_1 
\ast h_2.
\end{equation}
Therefore
\begin{equation}
K(x,y) = \int_{\rz^n} dz  L^1(x,z) L^2(z,y),
\end{equation}
with
\begin{equation}
 L^1(x,z)= f_1(x) h_1(x-z) \quad  L^2(z,y)= h_2(z-y) f_2(y).
\end{equation}
Observe 
\begin{equation}
\int_{\rz^n} \int_{\rz^n} dx dz| L^j(x,z)|^2 = \int_{\rz^n} \int_{\rz^n} dx dz |f_j(x)|^2 | h_j(z)|^2 
= \|f_j\|_2^2 \|\hat g\|_1 < \infty
\end{equation}
for $j=1,2$,
which implies the Lemma.

\end{proof}
\end{appendix}

\begin{thebibliography}{10}
\bibitem{ASS}
J. Avron, R. Seiler, B. Simon.
\newblock The Index of a Pair of Projectors.
\newblock {\em J. Funct. Anal.}, 120:220--237, 1994.
\bibitem{BG} A. Berthier, V. Georgescu.
\newblock On the Point Spectrum of Dirac operators.
\newblock {\em J. Funct. Anal.}, 71:309--338, 1987.

\bibitem{DES} J. Dolbeault, M.J. Esteban, E. S\'er\'e.
\newblock On the eigenvalues of operators with gaps. Application to Dirac operators.
\newblock {\em J. Funct. Anal.}, 174:208--226, 2000.

\bibitem{Dirac1934}
P.-A.-M. Dirac.
\newblock Th{\'e}orie du positron.
\newblock In Cockcroft, J.~Chadwick, F.~Joliot, J.~Joliot, N.~Bohr, G.~Gamov,
  P.A.M. Dirac, and W.~Heisenberg, editors, {\em Structure et
  propri{\'e}t{\'e}s des noyaux atomiques. Rapports et discussions du septieme
  conseil de physique tenu {\`a} Bruxelles du 22 au 29 octobre 1933 sous les
  auspices de l'institut international de physique Solvay. Publies par la
  commission administrative de l'institut.}, pages 203--212. Paris:
  Gauthier-Villars. XXV, 353 S., 1934.

\bibitem{Dirac1934D}
P.A.M. Dirac.
\newblock {Discussion of the infinite distribution of electrons in the theory
  of the positron.}
\newblock {\em Proc. Camb. Philos. Soc.}, 30:150--163, 1934.

\bibitem{Dyson1949}
F.~J. Dyson.
\newblock The radiation theories of {T}omonaga, {S}chwinger, and {F}eynman.
\newblock {\em Physical Rev. (2)}, 75:486--502, 1949.

\bibitem{Eff}
E. Effros. 
\newblock Why the circle is connected.
\newblock {\em Math. Intelligencer}, 11(1):27--35, 1989.



\bibitem{FrenchWeisskopf1949}
J.D. French and V.F. Weisskopf.
\newblock {The Electromagnetic Shift of Energy Levels.}
\newblock {\em Phys. Rev., II. Ser.}, 75:1240--1248, 1949.


\bibitem{GR}
W.~Greiner, J. Reinhardt.
\newblock {\em Quantum Electrodynamics}.
\newblock  Springer-Verlag, Berlin and Heidelberg
  and New York and Tokyo, 2 edition, 1994.

\bibitem{Greineretal1985}
W.~Greiner, B.~M{\"u}ller, and J.~Rafelski.
\newblock {\em Quantum Electrodynamics of Strong Fields}.
\newblock Texts and Mongraphs in Physics. Springer-Verlag, Berlin and Heidelberg
  and New York and Tokyo, 1 edition, 1985.
  
  
\bibitem{HS}
Ch. Hainzl, and H.~Siedentop.
\newblock {Non-perturbative Mass and Charge Renormalization in Relativistic No-photon QED.}
\newblock {\em Commun. Math. Phys}, (in press) arXiv: math-ph/0303043.

\bibitem{Heisenberg1934}
W.~Heisenberg.
\newblock {Bemerkungen zur {D}iracschen {T}heorie des {P}ositrons.}
\newblock {\em Z. Phys.}, 90:209--231, 1934.


\bibitem{JauchRohrlich1955}
J.~M. Jauch and F.~Rohrlich.
\newblock {\em The theory of photons and electrons. {T}he relativistic quantum
  field theory of charged particles with spin one-half}.
\newblock Addison-Wesley Publishing Company, Inc., Cambridge, Massachusetts,
  1955.

\bibitem{Kato1966}
Tosio Kato.
\newblock {\em Perturbation Theory for Linear Operators}, volume 132 of {\em
  Grundlehren der mathematischen {W}issenschaften}.
\newblock Springer-Verlag, Berlin, 1 edition, 1966.

\bibitem{KlausScharf1977T}
M.\ Klaus and G.\ Scharf.
\newblock The regular external field problem in quantum electrodynamics.
\newblock {\em Helv.\ Phys.\ Acta}, 50(6):779--802, 1977.

\bibitem{KlausScharf1977V}
M.\ Klaus and G.\ Scharf.
\newblock Vacuum polarization in {F}ock space.
\newblock {\em Helv.\ Phys.\ Acta}, 50(6):803--814, 1977.

\bibitem{KrollLamb1949}
Norman~M. Kroll and Willis~E. {Lamb~jun}.
\newblock {On the Self-Energy of a Bound Electron.}
\newblock {\em Phys. Rev., II. Ser.}, 75:388--398, 1949.


\bibitem{Milonni1994}
Peter~W. Milonni.
\newblock {\em The Quantum Vacuum: An Introduction to Quantum Electrodynamics}.
\newblock Academic Press, Inc., Boston, 1 edition, 1994.

\bibitem{NS}
G. Nenciu, G. Scharf.
\newblock On regular external fields in quantum electrodynamics.
\newblock {\em Helv.\ Phys.\ Acta}, 51: 412--424, 1977.
\bibitem{MPS}
P. J. Mohr, G. Plunien, G. Soff.
\newblock {QED corrections in Heavy Atoms.}
\newblock {\em Phys. Rep}, 293:227--369, 1998.

\bibitem{PauliRose1936}
W.~Pauli and M.E. Rose.
\newblock {Remarks on the Polarization Effects in the Positron Theory.}
\newblock {\em Phys. Rev., II. Ser.}, 49:462--465, 1936.


\bibitem{RMG}
J. Reinhardt, B. M\"uller, W. Greiner.
\newblock {Theory of positron production in heavy-ion collision.}
\newblock {\em Phys. Rev. A}, 24(1):103-128, 1981.

\bibitem{RS}
Michael Reed and Barry Simon.
\newblock {\em Methods of Modern Mathematical Physics}, volume 4: Analysis of
  Operators.
\newblock Academic Press, New York, 1 edition, 1978.

\bibitem{Schwinger1949}
Julian Schwinger.
\newblock {Quantum Electrodynamics II. Vacuum Polarization and Self-Energy.}
\newblock {\em Phys. Rev., II. Ser.}, 75:651--679, 1949.

\bibitem{Serber1935}
Robert Serber.
\newblock {Linear modifications in the Maxwell field equations.}
\newblock {\em Phys. Rev., II. Ser.}, 48:49--54, 1935.

\bibitem{Simon1979T}
Barry Simon.
\newblock {\em Trace Ideals and their Applications}, volume~35 of {\em London
  Mathematical Society Lecture Note Series}.
\newblock Cambridge University Press, Cambridge, 1979.


\bibitem{Uehling1935}
E.A. Uehling.
\newblock {Polarization effects in the positron theory.}
\newblock {\em Phys. Rev., II. Ser.}, 48:55--63, 1935.

\bibitem{Weid}
J. Weidmann.
\newblock {\em Linear Operators in Hilbert Spaces.}, 
\newblock Graduate Texts in Mathematics, 68. Springer-Verlag, New York-Berlin, 1980.

\bibitem{Weisskopf1936}
V.~Weisskopf.
\newblock {{\"U}ber die {E}lektrodynamik des {V}akuums auf {G}rund der
  {Q}uantentheorie des {E}lektrons.}
\newblock {\em {Math.-Fys. Medd., Danske Vid. Selsk.}}, 16(6):1--39, 1936.

\end{thebibliography}

\end{document}